\documentclass[a4paper]{spie}  
\usepackage[]{graphicx}

\title{Laser cooling of dense atomic gases by collisional redistribution of radiation and spectroscopy of molecular dimers in a dense buffer gas environment
} 

\author{Anne Sa{\ss}, Ralf Forge, Stavros Christopoulos, Katharina Knicker, Peter Moroshkin, and Martin Weitz\\
Institut f{\"u}r Angewandte Physik, Universit{\"a}t Bonn, Wegelerstr. 8, 53115 Bonn, Germany
}
\authorinfo{Corresponding author, A.S.: E-mail: sass@iap.uni-bonn.de}

\begin{document} 
  \maketitle 

\begin{abstract}
We study laser cooling of atomic gases by collisional redistribution of fluorescence. In a high pressure buffer gas regime, frequent collisions perturb the energy levels of alkali atoms, which allows for the absorption of a far red detuned irradiated laser beam. Subsequent spontaneous decay occurs close to the unperturbed resonance frequency, leading to a cooling of the dense gas mixture by redistribution of fluorescence. Thermal deflection spectroscopy indicates large relative temperature changes down to and even below room temperature starting from an initial cell temperature near \(700\,\rm K\). We are currently performing a detailed analysis of the temperature distribution in the cell. As we expect this cooling technique to work also for molecular-noble gas mixtures, we also present initial spectroscopic experiments on alkali-dimers in a dense buffer gas surrounding.  
\end{abstract}


\keywords{Laser cooling, spectroscopy in high buffer gas, atomic collisions}

\section{INTRODUCTION}
\label{sec:intro}  

\begin{figure}%
\centering
\includegraphics[width=0.42\textwidth]{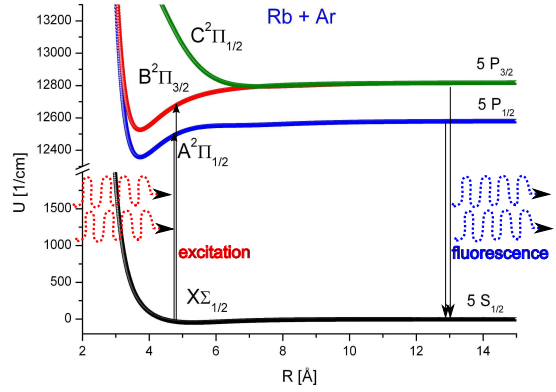}%
\caption{Cooling principle for the case of rubidium atoms subject to high pressure argon gas using calculated potential curves\cite{BerricheCo}. When an argon buffer gas atom collides with a rubidium atom, absorption of red detuned incident radiation becomes possible. The alkali atoms populate excited states and subsequent fluorescence originates close to the unperturbed resonances, thus having shorter wavelength than the incident laser beam and effectively cooling the sample.}%
\label{Coolprinciple}%
\end{figure}

Light as a tool for cooling matter was suggested for the first time already in 1929 by Peter Pringsheim \cite{Pringsheim}. During the last decades, various different approaches utilizing optical stimulation have produced fruitful results. Doppler cooling of dilute atomic gases\cite{HaenschSchawlow, ChuPhillips} is already a well-established technique, and anti-Stokes cooling in multilevel systems led to the optical refrigeration of solids\cite{DjeuWhitney, EpsteinCo, SeletskiyCo}. 
In this paper, we study the cooling of an ultradense alkali - noble gas mixture by collisional redistribution of radiation. This cooling mechanism was proposed in theoretical works by Berman and Stenholm \cite{BermanStenholm} almost 35 years ago. They discussed the effect of laser cooling and heating of an (atomic) two level system based on the energy loss during collisional aided excitation of atoms. Subsequent experiments at that time never reached the cooling regime \cite{GiacobinoCo}. 
Recently, our group succeeded in demonstrating relative cooling of dense alkali- noble gas mixtures using collisional redistribution of radiation \cite{VoglWeitz, AnneAPB, VoglJMO, VoglSpie}. 
Here, we report on current experiments using this technique. Also, we present initial spectroscopic experiments on an alkali-dimer - noble gas mixture which constitutes a promising candidate for redistributional laser cooling of molecules, following an alternative approach for the production of suitable molecules instead of laser ablation \cite{SimonSpie}. Our experiments are carried out using hot alkali atoms vapor in a noble gas environment under high pressure (up to \(230\,\rm bar\)). In this regime, the alkali resonances become strongly broadened so that their linewidth is of the order of the thermal energy \(k_{\rm B}T\) in frequency units\cite{VoglPRA}. The cooling mechanism can then be understood as follows: The strong, pressure induced, broadening of the lines allows for the absorption of a far red-detuned laser beam by rubidium atoms. The radiative lifetime of the rubidium electronically excited state exceeds the time of a collision by four orders of magnitude (\(27\,\rm ns\) compared to a few ps), allowing for the radiative deexcitation of the rubidium atoms, while at large distances from their collisional partners and thus closer to the unperturbed resonance frequency. Hence, the wavelength of the fluorescence is significantly shorter than the one of the absorbed light and energy extraction from the dense atomic ensemble becomes feasible, cooling the sample. A scheme of the cooling principle can be seen in Fig.\ref{Coolprinciple} for the case of rubidium atoms at the presence of high pressure argon noble gas. The here depicted potential curves of this Van-der-Waals-complex were taken from calculations of \(\rm[Rb^{+}]\) and \(\rm [e-Ar]\)-pseudopotentials \cite{BerricheCo}. 

\section{Redistribution of fluorescence}
\label{sec:setup}

\begin{figure}%
\centering
\includegraphics[width=0.5\textwidth]{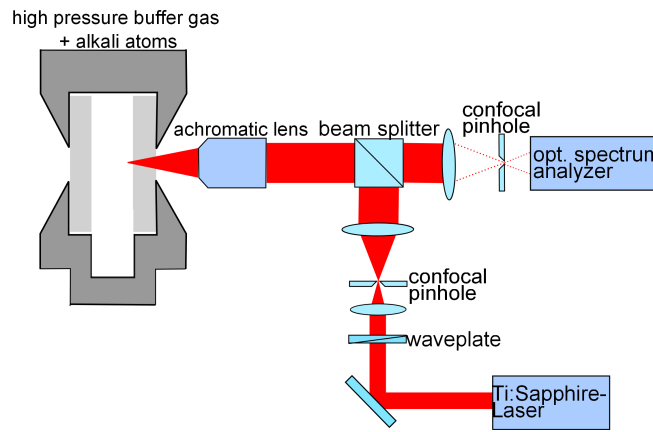}%
\caption{Confocal setup used to detect atomic fluorescence of the dense atomic mixture. The incident laser beam is spatially filtered by a confocal pinhole and afterwards focused into the heated cell. Atomic fluorescence is then collected in back direction, again spatially filtered and afterwards detected by an optical spectrum analyzer.}%
\label{ConfSetup}%
\end{figure}

This section offers insight into the experimental setup used to conduct redistribution of fluorescence measurements. The samples were prepared in a stainless-steel high-pressure cell with an inner volume of a few cubic centimeters, while optical access was provided by sapphire windows. Heating of the pressure cell is a prerequisite for the optical density necessary for the absorption of the cooling laser beam. For the experiments presented here, in order to achieve sufficient thermal alkali vapor pressure (\(n_{\rm Rb} \approx 10^{16}\,\rm cm^{-3}\)), the cell was heated up to \(700\,\rm K\) by an electrical heating system. The cell is also incorporated with a gas inlet, as well as a manometer, thus allowing for both constant monitoring of the pressure and conduction of pressure-dependent experiments. The confocal setup, as depicted in Fig.\ref{ConfSetup}, was used for the measurement of the fluorescent yield that allows for the determination of the redistribution of fluorescence, necessary for this particular cooling technique. 
Cooling radiation near the alkali D-resonances was generated with a continuous-wave Ti:Sapphire ring laser. The incident laser beam was guided through the cell volume using the cell's sapphire windows (\(18\,\rm mm\) diameter). All experiments were performed with noble gases of high purity to avoid quenching effects. To suppress back reflections, e.g. from the cell's front window, both the incident laser beam and the emitted fluorescence were spatially filtered with confocal pinholes. The incident laser beam was focussed to a beam diameter of approximately \(10\,\rm \mu m\), and fluorescence was collected from that area and subsequently measured by an optical spectrum analyzer.

\begin{figure}%
\centering
\includegraphics[width=0.5\textwidth]{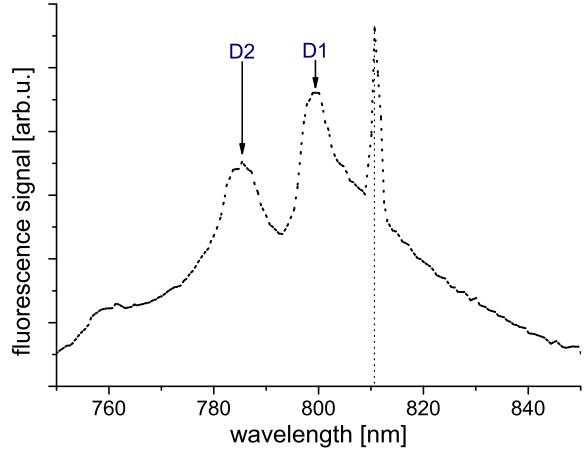}%
\caption{Recorded fluorescence of rubidium atoms subject to \(160\,\rm bar\) argon buffer gas at a temperature \(T= 580\,\rm K\). Both the \(D1\) and \(D2\) - transitions are spectrally resolved. The vertical dashed line shows the wavelength of the incident laser beam along with residual scattered laser beam radiation.}%
\label{FigFluo}%
\end{figure}

A typical fluorescence spectrum can be seen in Fig.\ref{FigFluo}. Laser radiation at \(\lambda \approx 810\,\rm nm\) was irradiated into the cell filled with rubidium atoms subject to \(160\,\rm bar\) argon gas. Here, the cell was heated up to a temperature \(T = 580\,\rm K\). A clear redistribution towards the centre of the D-resonances can be observed. The spectrum also shows a peak at the incident laser wavelength which is due to remaining scattered laser beam radiation. The expected cooling power \(P_{\rm cool}\) can be calculated from the average frequency \(\nu_{\rm fl}\) of the fluorescence from the corresponding spectrum and also using the absorption probability \(a(\nu)\), incident laser frequency \(\nu\) and power \(P_{\rm opt}\):
\begin{equation}
P_{\rm cool} = P_{\rm opt}a(\nu) \frac{\nu_{\rm fl}-\nu}{\nu}.
\label{eqCoolPower}
\end{equation}
Under typical experimental conditions, the expected cooling power reaches around \(100\,\rm mW\).

\section{Redistribution laser cooling and thermal deflection spectroscopy}
\label{sec:coolatom}
\begin{figure}%
\centering
\includegraphics[width=0.5\columnwidth]{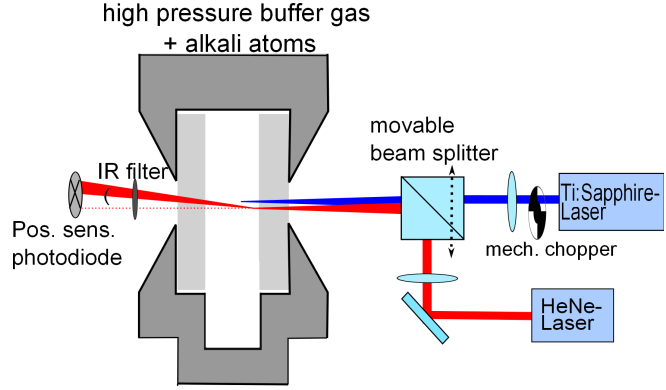}%
\caption{Sketch of the thermal deflection spectroscopy setup to monitor the induced temperature distribution. Collinearly, the cooling laser beam and the helium-neon-laser beam are guided into the high pressure cell. The lateral beam displacement can be varied, therefore the angular deflection can be measured position-depended by a position sensitive photodiode.}%
\label{tdssetup}%
\end{figure}

To measure the temperature distribution of the gas mixture induced by the cooling laser beam, thermal deflection spectroscopy (TDS) is applied\cite{JacksonCo, SpearCo}. In this technique, a second, non-resonant He-Ne-probe beam is directed into the cell, collinear to the cooling laser path. The lateral distance between the two laser beams can be varied using a motorized translation stage. The temperature gradient induced by the cooling laser causes a change in density and therefore a spatial variation of the refractive index of the gas, resulting in a prismatic deflection of the probe beam. While periodically blocking the cooling laser beam with a mechanical shutter, the induced angular deflection of the probe beam can be measured by a position sensitive photodiode placed behind the cell. By scanning the position of the probe laser beam with respect to the cooling laser beam, the temperature distribution inside the cell can be mapped. In the experiments presented here, the lateral offset between the two laser beams was varied in steps of \(100\,\rm \mu m\) over a range of a few millimetres. The cooling and the probe laser beams were focused to focal diameters of \(400\,\rm \mu m\) (cooling laser) and \(200\,\rm \mu m\) (probe beam) respectively. Fig.\ref{tdssetup} shows the experimental setup for the cooling measurements. 
The temperature distribution can be derived from the measured deflection angle of the probe beam. When considering the propagation of a laser beam with a Gaussian intensity distribution through a spatially variable refraction index profile, the deflection angle \(\phi\) as a function of the change towards the \(z\)-direction can be expressed as follows:
\begin{equation}
\phi = \frac{1}{n_0}\int{dz \frac{dn}{dT}\frac{dT}{dr}},
\label{eq1}
\end{equation}
 where \(n_0\) is the refractive index of the (noble) gas under normal conditions. As experiments are performed with a non-resonant probe beam, \(dn/dT\) can be simplified using the  Lorentz-Lorenz-relation\cite{BornWolf}, and also for the sake of simplicity assuming an ideal gas and small refractive index changes:
 \begin{equation}
\frac{dn}{dT} \approx -\frac{(n-1)}{T} .
\label{eq2}
\end{equation} 
 The distribution of the temperature change \(\Delta T(r, z)\) can then be written as:
 \begin{equation}
\Delta T(r,z) = \frac{T}{(n-1)}\cdot \frac{\alpha \cdot e^{-\alpha z}}{(1-e^{-\alpha L})}\int_r{dr' \phi(r')} ,
\label{eq3}
\end{equation}
where \(r\) is the transverse displacement, \(L\) the inner length of the cell, \(T\) the initial temperature of the gas and \(n\) the refractive index at experimental conditions. We assume that the Gaussian laser beam is removing thermal energy with a rate following its intensity distribution. In our present model, heat transport is considered to occur exclusively in the radial direction because the absorption length, \(1/\alpha\), is much longer than the laser beam focal radius \(w_{\rm 0}\). The heat transfer can be described by 
\begin{equation}
\nabla \cdot (\kappa \nabla T) = -Q(r,z),
\label{eq5}
\end{equation}  
where \(\kappa\) is the thermal conductivity of the buffer gas. The cooling source \(Q(r, z)\) of a Gaussian laser beam can be expressed \cite{GordonCo, ZhuCo}:
\begin{equation}
Q(r,z) \propto \alpha \cdot P_{\rm cool} \cdot \exp(-2r^2/w_0^2) \cdot \exp(-\alpha z) .
\label{eq4}
\end{equation}
where \(P_{\rm cool}\) denotes the cooling power and \(\alpha\) the absorption coefficient. 
\begin{figure}%
\centering
\includegraphics[width=0.5\columnwidth]{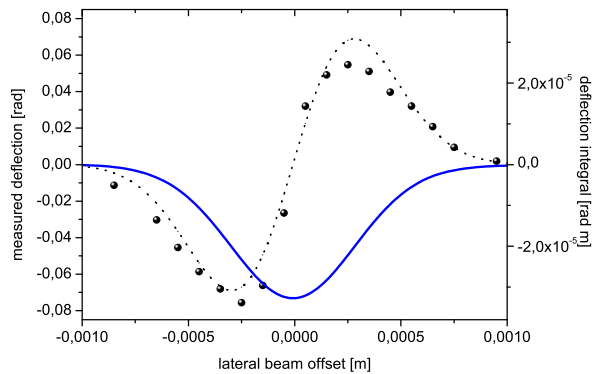}%
\caption{Measured deflection angle versus different transverse displacements between cooling and probe laser beam for rubidium atoms subject to \(220\,\rm bar\) argon gas. The black dots indicate the measured deflection, while the dashed line represents a fit to the data. The solid line gives the transverse integral of the deflection angle, from which the temperature drop in the laser beam focus can be calculated.}%
\label{coolRbAr}%
\end{figure}
Fig.\ref{coolRbAr} shows a typical deflection measurement for the case of rubidium atoms subject to \(220\,\rm bar\) argon buffer gas at an initial temperature of \(T= 680\,\rm K\). \(2\,\rm W\) incident power of Ti:Sapphire-laser radiation at a wavelength of \(\lambda = 810\,\rm nm\) was irradiated into the high pressure cell. The angular deflection was measured at different beam displacements and afterwards, the integral of \(\int_r{dr' \phi(r')}\) was calculated. Under the above assumptions, the depicted distribution can be related to a maximum temperature change of \(\Delta T \approx - 490\,\rm K\) in the laser beam focus, almost one order of magnitude larger than in previously reported results \cite{VoglWeitz, VoglJMO} for rubidium atoms subject to high pressure argon. We attribute this mainly to the higher cooling laser intensity due to a smaller beam focus (\(w_{\rm 0} \approx 200\,\rm \mu m\) for the data set of Fig.\ref{coolRbAr}), as well as to a higher optical density at the used hot cell temperature.  
The model that we presently use to calculate the temperature changes from the experimental deflection data however neglects a possible variation of the absorption coefficient \(\alpha\) with temperature, whose possible implications need to be accounted for temperature changes that are of the same order as the absolute temperature. The quoted preliminary value for the temperature drop assumes that the absorption coefficient is constant. We are also currently carrying out more detailed heat transfer calculations. Inspection of equation \ref{eq5} reveals that the temperature drop achieved by the cooling should increase using a gas with lower thermal conductivity. Thus, use of heavier noble gases, such as krypton and xenon, should allow for larger temperature changes in comparison with the presently used argon buffer gas. Note that the collisional cooling technique requires highly purified gases (we presently use noble gas with a specified impurity level below \(10^{-5} - 10^{-6}\)) to reduce nonradiative quenching. The availability of such pure gas samples of the heavier gases can be problematic. Using commercially available gases, we have nevertheless achieved encouraging first cooling results both with rubidium-krypton and rubidium-xenon gas mixtures.

\section{Spectroscopy of a dense molecular-noble gas mixture}
\label{sec:specmol}

\begin{figure}[h]%
\centering
\includegraphics[width=0.5\columnwidth]{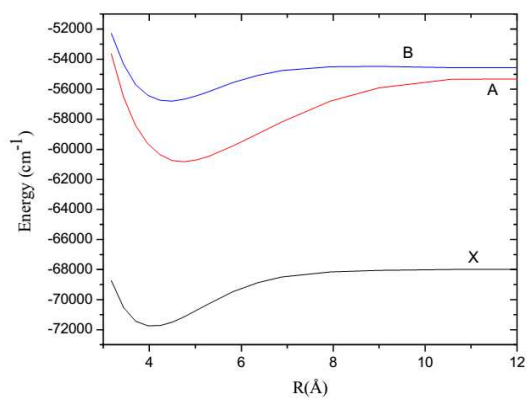}%
\caption{Calculated\cite{SpiegelmannCo} energy of the ground state X (black line), the first excited state A (red line) and the second excited state B (blue line) of the rubidium diatomic molecule (\(\rm Rb_{\rm 2}\)) as a function of the distance between the two rubidium atoms of the molecule.}%
\label{RbMolPot}%
\end{figure}

The process of redistribution laser cooling is also expected to be observed for a molecular-noble gas mixture. Towards this goal, we performed several experiments with alkali dimers in a high pressure argon environment. Alkali dimers, namely molecules consisting of two identical alkali atoms bound in a potential well, are expected to exist in a small percentage, namely \(3\%\) when compared to the monomers, in the hot alkali vapor\cite{GuptaCo} produced in our experiments. Their presence can be subsequently easily verified via absorption measurements. We expect the redistribution process to be similar to that observed for the atomic case. In addition to the kinetic and electronic energy, the molecule can save energy in rotations and vibrations. The calculated\cite{SpiegelmannCo} potential curves of the ground (X) and first two excited states (A, B) of the rubidium dimer are depicted in Figure \ref{RbMolPot} as a function of the interatomic distance. 

\begin{figure}%
\centering
\includegraphics[width=0.5\columnwidth]{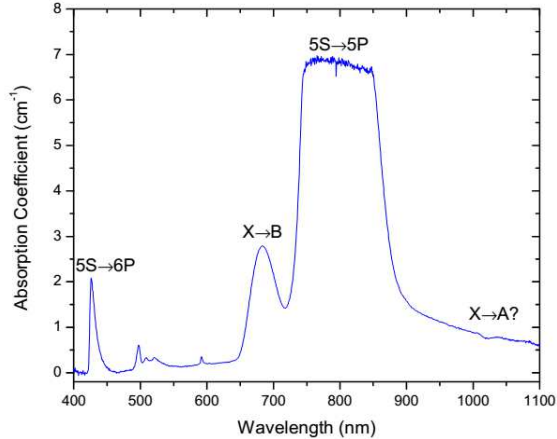}%
\caption{Absorption coefficient of rubidium vapor subject to \(180\,\rm bar\) argon buffer gas at a temperature of \(T = 670\,\rm K\), as measured by spectrally analyzing the transmission of a rubidium-buffer gas cell under irradiation by a spectrally broadband halogen lamp. The notes above the peaks indicate the observed atomic and molecular absorption bands, referring to \cite{VdovicCo}.}%
\label{AbsDimer}%
\end{figure}

In the process of redistribution of fluorescence, the laser photon is absorbed by the alkali molecule with the synergy of the collision partners, and is subsequently excited to an electronic state. However, beside the electronic state, a change of the rotational and vibrational state is also possible. While the rotational quantum number can only be changed by 1, due to angular momentum conservation, the change of the vibrational state is not limited. When the laser photon is absorbed, the state is changed vertically in the level scheme, thus the change of the vibrational state may become large, depending on the horizontal shift of the
minima of the two states. After the absorption, the excited state is expected to thermalize due to collisions with buffer gas atoms, therefore the vibrational and rotational structure is expected to thermalize quickly. This thermalization can either heat or cool the sample. Analogous to the atomic case, the efficiency of the redistribution process is expected to depend on several parameters. The detuning should be in the order of magnitude of the kinetic energy of the molecules, \(k_{\rm B}T\). Furthermore, in order to achieve an adequate number of collisions, both the pressure and temperature need to be high. More particularly, due to the fact that the ratio between alkali dimers and monomers is an exponential function of temperature, it is important to work at sufficiently high temperatures. Finally, a high number of excitation events can be achieved using high laser intensities. \\
The redistribution process is favoured if the detuned laser light matches the energy difference for certain values of the distance between alkali molecule and noble gas atom. Since this three body problem cannot be solved analytically and has not been investigated yet, there are no known potential lines for this problem as for the atomic case (Rb+Ar)\cite{BerricheCo}.\\ 
In order to investigate the presence of alkali dimers in dense buffer gas, absorption spectroscopy is performed using an ordinary halogen lamp. The light passes through the cell and is subsequently collected by an optical spectrum analyzer. 
We measure absorption spectra for a mixture of rubidium and \(180\,\rm bar\) argon at a temperature of \(T = 670\,\rm K\). The absorption versus the wavelength, as derived from the transmission spectrum is shown in Fig.\ref{AbsDimer}. 
The largest peak of the spectrum arises from the atomic D-line transitions. Due to the high pressure, these transitions are strongly broadened and cannot be spectrally resolved. The maximum shown value for the absorption is limited by the sensitivity of the spectrometer and is smaller than the expected absorption at these resonances. A further absorption peak can be seen at the spectral position of \(\lambda \approx 422\,\rm nm\), which we attribute to the atomic (\(5S \rightarrow 6P\)) - resonance. Due to the pressure broadening being much larger than the fine structure splitting of the 6P state, the corresponding substructure is not resolved.\\ Furthermore, the absorption band of the (\(X \rightarrow B\)) - molecular transition in the range of \((650 - 730)\,\rm nm\) can be easily observed, in agreement with earlier experiments\cite{GuptaCo, VdovicCo} in the dilute gas limit. In the range of \((800 - 1100)\,\rm nm\), the presence of the weaker (\(X \rightarrow A\)) - absorption band cannot be resolved, most probably due to the low-frequency wing of the strong atomic D-line transitions which here still dominates. The origin of the absorption peaks in the green has to be investigated further.
In conclusion, characteristic transitions for atomic and molecular rubidium have been observed. The presence of \(\rm Rb_2\) in our high pressure buffer gas sample is thus established. Preliminary experiments with a diode laser at a fixed wavelength \(\lambda = 920\,\rm nm\) indicate the existence of a weak fluorescence band in the range within \((800 - 1100)\,\rm nm\), also known from \cite{BeucCo}. Here, further investigations with a widely tunable laser source are needed to check for clear redistribution of fluorescence.

\section{Conclusion}
\label{sec:conclusion}
We investigated the effect of redistributional laser cooling of dense alkali-noble gas mixtures. The irradiation with the cooling laser results in large angular deflection of a probe laser beam from the induced thermal lens. We are presently in the process of obtaining the corresponding temperature change from the cooling. Future experiments will include the study of homogeneous nucleation in saturated vapor and the application of this cooling mechanism on molecular gases subject to high pressure buffer gas. Towards this goal, we here presented initial spectroscopic investigations on the rubidium dimer in high pressure argon environment.

\acknowledgments     
 
This work was supported by the Deutsche Forschungsgemeinschaft, grant No: We 1748-15.



\end{document}